\documentclass[graybox]{svmult}

\usepackage{type1cm}        
\usepackage{makeidx}         

\usepackage{graphicx}        

\usepackage{multicol}        
\usepackage[bottom]{footmisc}

\usepackage{newtxtext}       %
\usepackage[varvw]{newtxmath}       

\makeindex             

\theoremstyle{Axiom}
\newtheorem{axiom}{Axiom}

\begin{document}

\title*{Topological Data Analysis for Directed Dependence Networks of Multivariate Time Series Data}
\author{Anass B. El-Yaagoubi and Hernando Ombao}
\institute{Anass B. El-Yaagoubi \at King Abdullah University of Science and Technology, Thuwal 23955-6900, Kingdom of Saudi Arabia, \email{anass.bourakna@kaust.edu.sa}.
\and Hernando Ombao \at King Abdullah University of Science and Technology, Thuwal 23955-6900, Kingdom of Saudi Arabia, \email{hernando.ombao@kaust.edu.sa}.}
%
%
\maketitle

\abstract*{Topological data analysis (TDA) approaches are becoming increasingly popular for studying the dependence patterns in multivariate time series data. In particular, various dependence patterns in brain networks may be linked to specific tasks and cognitive processes, which can be altered by various neurological impairments such as epileptic seizures. Existing TDA approaches rely on the notion of distance between data points that is symmetric by definition for building graph filtrations. For brain dependence networks, this is a major limitation that constrains practitioners to using only symmetric dependence measures, such as correlations or coherence. However, it is known that the brain dependence network may be very complex and can contain a directed flow of information from one brain region to another. Such dependence networks are usually captured by more advanced measures of dependence such as partial directed coherence, which is a Granger causality based dependence measure. These dependence measures will result in a non-symmetric distance function, especially during epileptic seizures. In this paper we propose to solve this limitation by decomposing the weighted connectivity network into its symmetric and anti-symmetric components using matrix decomposition and comparing the anti-symmetric component prior to and post seizure. Our analysis of epileptic seizure EEG data shows promising results.}

\abstract{Topological data analysis (TDA) approaches are becoming increasingly popular for studying the dependence patterns in multivariate time series data. In particular, various dependence patterns in brain networks may be linked to specific tasks and cognitive processes, which can be altered by various neurological impairments such as epileptic seizures. Existing TDA approaches rely on the notion of distance between data points that is symmetric by definition for building graph filtrations. For brain dependence networks, this is a major limitation that constrains practitioners to using only symmetric dependence measures, such as correlations or coherence. However, it is known that the brain dependence network may be very complex and can contain a directed flow of information from one brain region to another. Such dependence networks are usually captured by more advanced measures of dependence such as partial directed coherence, which is a Granger causality based dependence measure. These dependence measures will result in a non-symmetric distance function, especially during epileptic seizures. In this paper we propose to solve this limitation by decomposing the weighted connectivity network into its symmetric and anti-symmetric components using matrix decomposition and comparing the anti-symmetric component prior to and post seizure. Our analysis of epileptic seizure EEG data shows promising results.}

\section{Introduction}
\label{sec:introduction}

Over the past twenty years, topological data analysis (TDA) has witnessed significant advances that aim to study and understand various patterns in the data, thanks to the pioneering work of \cite{BARCODES_FIRST}, \cite{EDELSBRUNNER_HARER}, \cite{TDA_EDELSBRUNNER} and \cite{BARCODES}.  Many TDA techniques for the analysis of various types of data have emerged, like barcodes, persistence diagrams and persistence landscapes. These TDA features aim to summarize the intrinsic shape of the data (cloud of points or weighted network usually), by keeping track of the specific scales at which any topological feature (e.g., holes and cavities etc.) is either born or dead. The goal of such approaches is to provide insight on the geometrical patterns included in high dimensional data that are not visible using conventional approaches.

There has been an increasing trend in the literature that applies the TDA techniques to data sets with a temporal structure (e.g., financial time series, brain signals etc.). For such data sets, it has been proposed over the years that dependence networks of multivariate time series data, particularly for multivariate brain signals such as electroencephalograms (EEG) and local field potentials (LFP), can provide promising results, as shown in \cite{TDA_MULTIVARIATE_TS_ANASS}. 

The topology of the structural and functional brain networks is organized according to principles that maximize the (directed-)flow of information and minimize the energy cost for maintaining the entire network, such as small world networks, see \cite{SMALL_WORLD_NETWORKS}, \cite{BRAIN_NETWORKS_ORGANIZATION} and \cite{STRUCTURE_FUNCTION_BRAIN_NETWORKS}. More formally, define $X(t) \in \mathbb{R}^{d}$ to be a vector-valued stochastic brain process at time $t$. Due to physiological reasons, the influence of $X_q(t)$ on $X_p(t)$ may differ from the influence of $X_p(t)$ on $X_q(t)$, where $X_p(t)$ and $X_q(t)$ are two univariate time series components of $X(t)$.

Neurological disorders, such as Alzheimer's disease, Parkinson's disease and Epilepsy may alter the topological structure of the brain network. For instance, it is known that the onset of epileptic seizures can alter brain networks and, in particular, brain connectivity, leading to multiple studies that compare the pre-ictal, inter-ictal and post-ictal network characteristics. However, studying this change in the topology of the brain network based on a symmetric measure of dependence such as correlation or coherence is going to result in the loss of information regarding the spatio-temporal evolution of the seizure process (how abnormal electrical activity in one brain region may spread to other regions). Indeed, it is know that brain seizures are usually initiated from a localized region (or multiple sub-regions), which then propagate to the rest of the network, for more details refer to \cite[Chapter~1]{EPILEPSY_INTRO}. Such propagation ought to be non-symmetric, as there is directed flow of information going from the source region to the rest of the brain.

Therefore, in order to capture this asymmetry in the brain connectivity we ought to use a non-symmetric measure of dependence, that allows for the influence of a channel $X_q(t)$ on another channel $X_p(t)$ to be different from the reverse influence, i.e., of channel $X_p(t)$ on channel $X_q(t)$. In particular, partial directed coherence (PDC), which estimates the intensity of information flow from one brain region to another based on the notion of the Granger causality, i.e., the ability of time series components to predict each other at various lag values. Refer to Section \ref{sec:VAR_PDC} for more details regarding PDC. Using the distance measure based on PDC in \cite{TDA_MULTIVARIATE_TS_ANASS} introduces the following problem. For a function $d$ to be a valid distance function it has to respect the four axioms of a distance:
\begin{axiom}
    $d(x, y) \geq 0$
    \label{axiom_1}
\end{axiom}
\begin{axiom}
    $d(x, y) = 0 \iff x = y$
    \label{axiom_2}
\end{axiom}
\begin{axiom}
    $d(x, y) = d(y, x)$
    \label{axiom_3}
\end{axiom}
\begin{axiom}
    $d(x, y) \leq d(x, z) + d(z, y)$
    \label{axiom_4}
\end{axiom}
After defining the distance uding PDC (e.g., $d=1-PDC$), we notice immediately that it fails to satisfy Axiom \ref{axiom_3}, i.e., $d(x, y) \neq d(y, x)$ which is a significant restriction. Some of the aforementioned axioms (such as the second and the fourth) can be relaxed in some situations. However, Axiom \ref{axiom_3} cannot be ignored, and doing so will lead to major conflicts and contradictions when creating the Rips-Vietoris filtration.

Motivated by this challenge, we propose a novel approach that will allow existing TDA techniques to assess the topology of the asymmetry in oriented brain networks. We propose an approach based on the decomposition of a weighted network into its symmetric and anti-symmetric components. In Section 2, we provide a brief review of vector autoregressive (VAR) models and PDC. In Section 3, we present the network decomposition approach. In Section 4, we provide a brief review of persistent homology and Vietoris-Rips filtration. In Section 5, we analyze the impact of an epileptic seizure on the flow of information within the brain using an EEG data set.

\section{VAR models and PDC}
\label{sec:VAR_PDC}

One standard approach to modeling brain signals is to use parametric models such as VAR models, see \cite{VAR_1}, \cite{VAR_2}, \cite{VAR_4} and \cite{VAR_3}. This class of models is very flexible, and has the ability to capture all linear dependencies in the multivariate time series, see \cite{LUTKEPOHL}. Given a stationary multivariate time series $X(t)\in \mathbb{R}^d$, the VAR model of order $K$ is expressed in the following way:
\begin{align}
    X(t) &= \sum_{k = 1}^{K} \Phi_{k} X(t-k) + E(t), \\
    E(t) &\sim \mathcal{N}(0, \Sigma_E),
\end{align}
where $X(t)$ is $d$-dimensional vector of observations, and $\Phi_{k}$ is $d\times d$ mixing weight matrix for lag $k$, which are chosen such that $X(t)$ is causal, $E(t)$ is the innovations vector. The model order $K$ is usually selected based on various information criterion, such as Akaike
information criterion (AIC) and Bayesian information criterion (BIC), see \cite{AIC}, \cite{BIC} and \cite{TSA_SHUMWAY_STOFFER}.

The concept of PDC, as presented in \cite{Bacala_Connectivity}, is supposed to represent the concept of Granger causality in the frequency domain, as it measures the intensity of information flow between a pair of EEG channels. Assuming we fit a VAR model to the observed multivariate brain signals, we get the following expression for the Fourier transform of the VAR model parameters:
\begin{align}
    \overline{A}(\omega) &= I - \sum_{k=1}^K \Phi_k \exp{(-i2\pi k \omega)},
\end{align}
which in turn is used as follows to compute the PDC:
\begin{align}
    PDC_{p, q}(\omega) &= \frac{\big|\overline{A}_{p, q}(\omega)\big|}{\sqrt{\overline{A}_{., q}^H(\omega) \overline{A}_{., q}(\omega)}},
\end{align}
where $PDC_{p, q}(\omega)$ denotes the direction and intensity of the information flow from channel $q$ to channel $p$ at the frequency $\omega$ and the symbol $(.)^H$ denotes the Hermitian transpose, when the dimension $d$ is equal to one then the Hermitian transpose becomes the complex conjugate. From this definition, it is obvious that PDC is not symmetric measure of information as $PDC_{p, q}(\omega) \neq PDC_{q, p}(\omega)$. Such a PDC cannot be used directly with TDA to analyze the shape of the network. Therefore, another approach is necessary,  which we will develop in the next section.

\begin{figure}
\centering
\includegraphics[width=.9\linewidth]{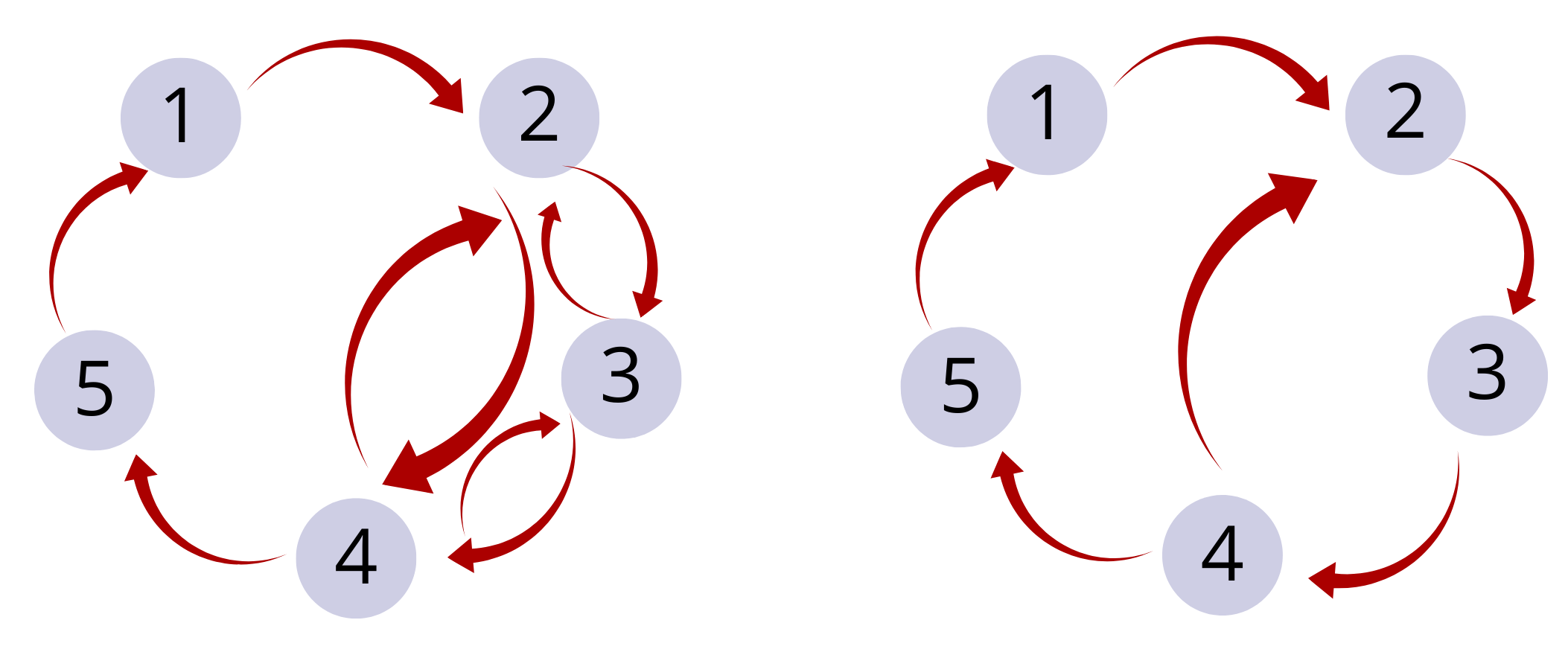}
\caption{Example of distinct directed dependence networks. Symmetric and non-symmetric cyclic structure (Left) and non-symmetric only cyclic structure (Right).}
\label{fig:two_different_directed_dependence_networks}
\end{figure}

Assume we observe two multivariate processes $Y^{(1)}(t)$ and $Y^{(2)}(t)$ that are mixtures of AR(2) processes $Z^{(k)}_j(t)$ as follows:

\begin{align*}
    Y_1^{(1)}(t) &= Y_5^{(1)}(t-1) + Z^{(1)}_1(t), \\
    Y_2^{(1)}(t) &= Y_1^{(1)}(t-1) + Y^{(1)}_4(t-1) + Z^{(1)}_2(t), \\
    Y_3^{(1)}(t) &= Y_2^{(1)}(t-1) + Y^{(1)}_4(t-1) + Z^{(1)}_3(t), \\
    Y_4^{(1)}(t) &= Y_2^{(1)}(t-1) + Y^{(1)}_3(t-1) + Z^{(1)}_4(t), \\
    Y_5^{(1)}(t) &= Y_4^{(1)}(t-1) + Z^{(1)}_5(t),
\end{align*}
\begin{align*}
    Y_1^{(2)}(t) &= Y_5^{(2)}(t-1) + Z^{(2)}_1(t), \\
    Y_2^{(2)}(t) &= Y_1^{(2)}(t-1) + Y_4^{(1)}(t-1) + Z^{(2)}_2(t), \\
    Y_3^{(2)}(t) &= Y_2^{(2)}(t-1) + Z^{(2)}_3(t), \\
    Y_4^{(2)}(t) &= Y_3^{(2)}(t-1) + Z^{(2)}_4(t), \\
    Y_5^{(2)}(t) &= Y_4^{(2)}(t-1) + Z^{(2)}_5(t),
\end{align*}
with distinct lead-lag directed dependence networks, as can be seen in Figure \ref{fig:two_different_directed_dependence_networks}. Choosing $Z^{(k)}_j(t)$ to be AR(2) processes allows this dependency pattern to be frequency specific. Both networks are oriented and have multiple edges in common; however, they present distinct oriented connectivity among the nodes $2-4$.

\begin{figure}
    \centering
    \includegraphics[width=.9\linewidth]{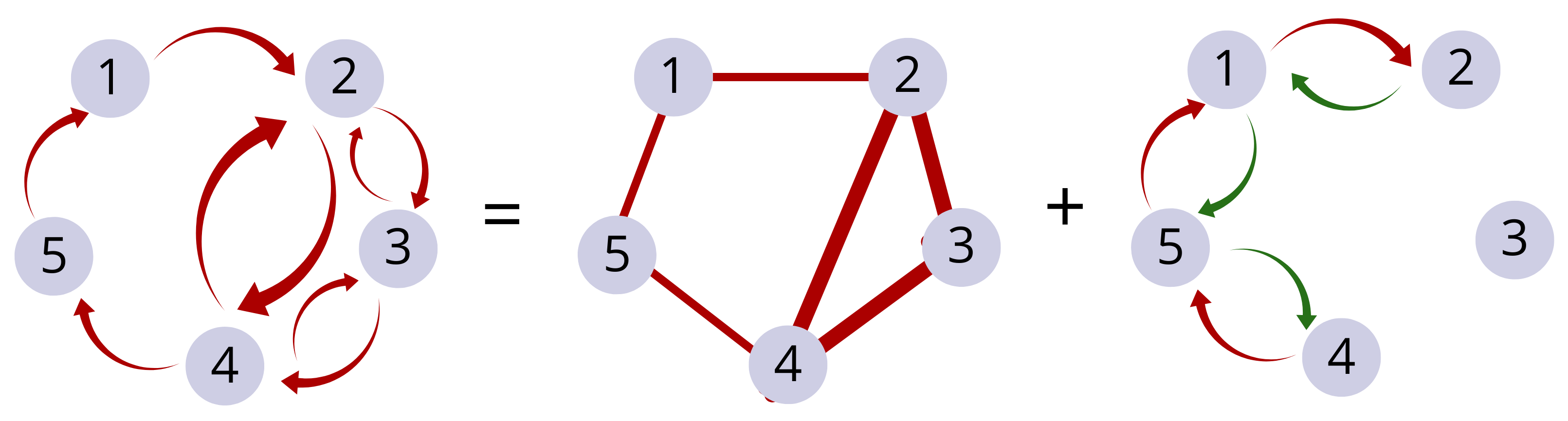}
    \caption{First example of non-symmetric network decomposition. Original network (Left), symmetric component (Middle) and anti-symmetric component (Right).}
    \label{fig:first_network_decomposition}
\end{figure}

\begin{figure}
    \centering
    \includegraphics[width=.9\linewidth]{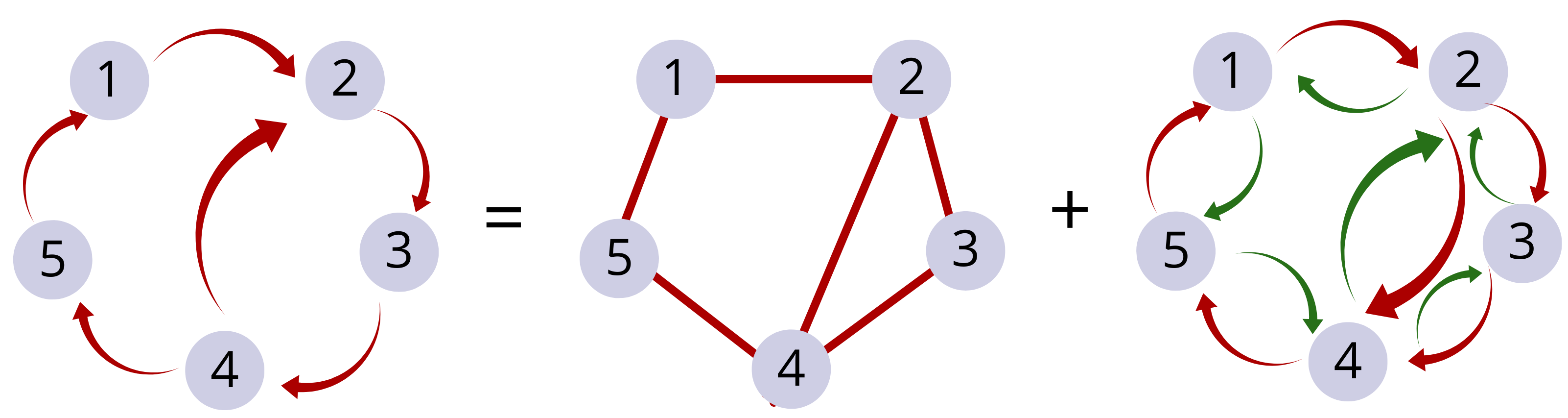}
    \caption{Second example of non-symmetric network decomposition. Original network (Left), symmetric component (Middle) and anti-symmetric component (Right).}
    \label{fig:second_network_decomposition}
\end{figure}

Using a non-oriented measure of dependence such as coherence or correlation would result in misleading conclusions, as these measures will ignore the orientation of the dependence structure. Hence, the two distinct structures will be confused into the same structure. However, using an oriented measure of dependence such as PDC will fit the structure properly. Therefore, the latter approach will enable us to take into account the orientation in the dependence structure by decomposing oriented networks into their symmetric and anti-symmetric components as can be seen in Figures \ref{fig:first_network_decomposition} and \ref{fig:second_network_decomposition}.

From Figures \ref{fig:first_network_decomposition} and \ref{fig:second_network_decomposition}, it is clear that applying TDA to the symmetric component (Middle) will results in disastrous conclusion, as it will not be able to distinguish between the two networks, since they share very similar topological structure. However, applying TDA to the anti-symmetric component (Right) will capture the topological differences between the two networks, as the first example doesn't contain any asymmetric cycles, whereas the second example contains two main cycles.

\section{Network decomposition}
\label{sec:network_decomposition}

\begin{figure}
    \centering
    \includegraphics[width=.9\linewidth]{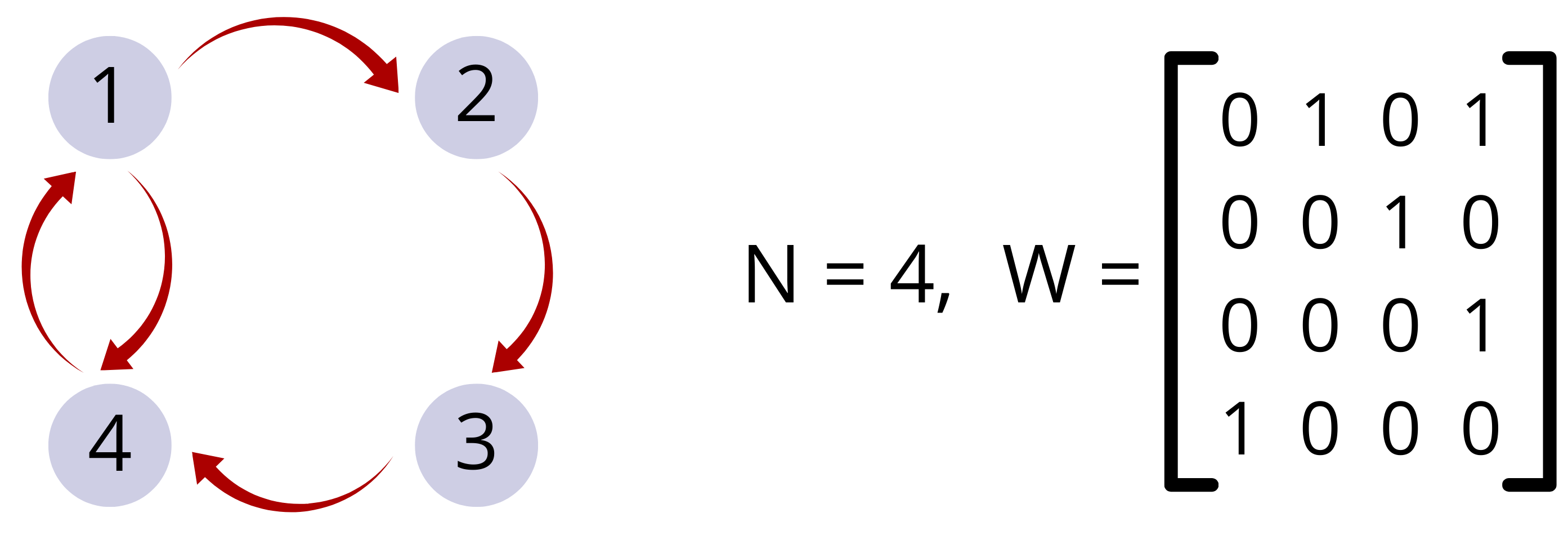}
    \caption{Original weight matrix and graph representation.}
    \label{fig:weight_matrix_visualization}
\end{figure}

\begin{figure}
    \centering
    \includegraphics[width=.9\linewidth]{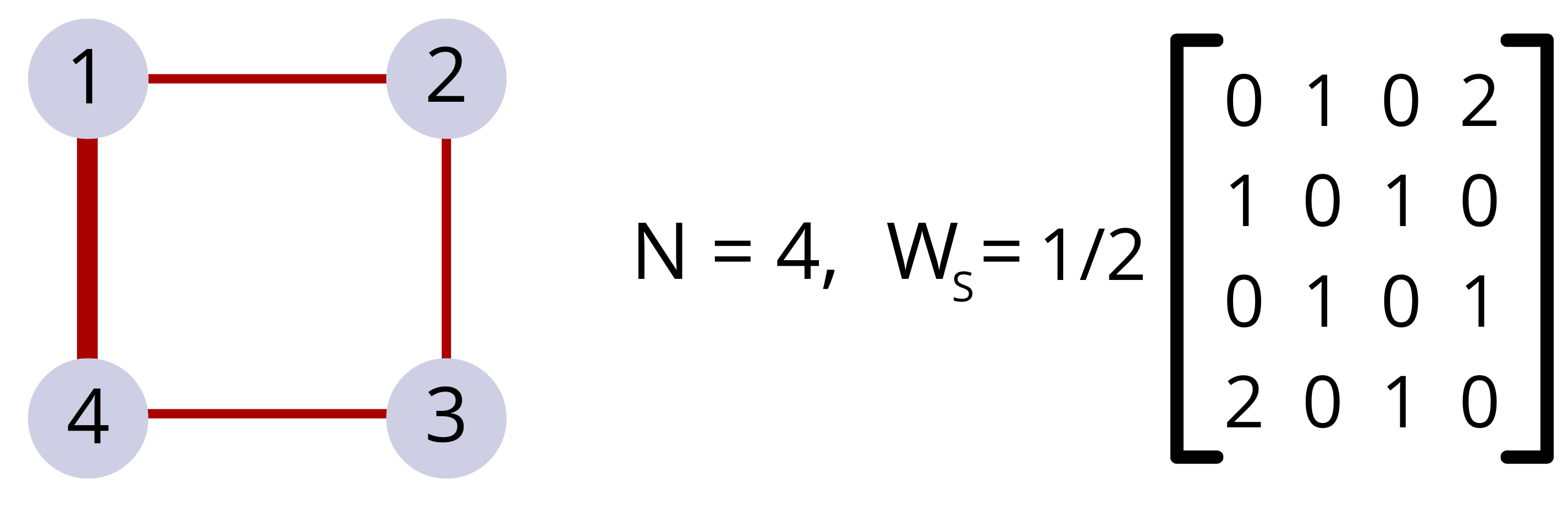}
    \caption{Symmetric component representation.}
    \label{fig:weight_matrix_visualization_symmetric}
\end{figure}

\begin{figure}
    \centering
    \includegraphics[width=.9\linewidth]{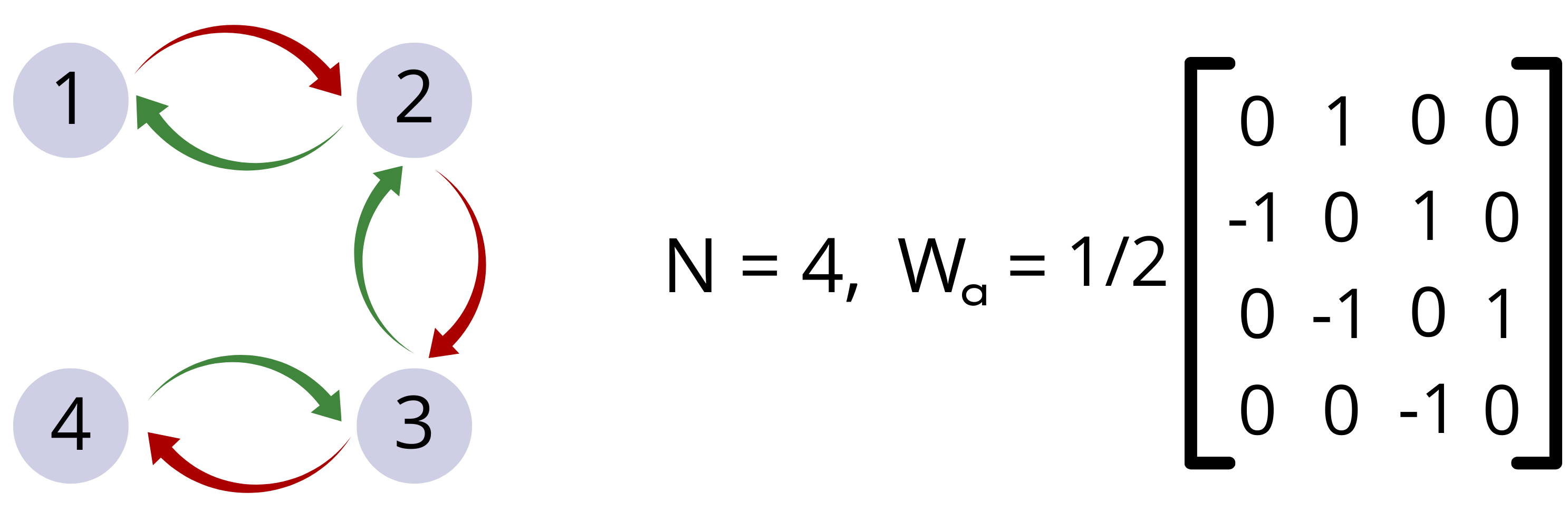}
    \caption{Anti-symmetric component representation.}
    \label{fig:weight_matrix_visualization_anti_symmetric}
\end{figure}

There are multiple techniques for decomposing weighted network into multiple networks with various properties. In this section, we consider the transformation that decomposes a network into a unique pair of symmetric and anti-symmetric components. Given an oriented weighted network $G = (N, W)$, see Figure \ref{fig:weight_matrix_visualization}, with $|N|$ nodes and weight matrix $W$, we propose the following graph decomposition that results in two graphs $G_s$ and $G_a$,  where $G_s = (N, W_s)$ is the symmetric component and $G_a = (N, W_a)$ is the anti-symmetric component as can be seen in Figures \ref{fig:weight_matrix_visualization_symmetric} and \ref{fig:weight_matrix_visualization_anti_symmetric}.
\begin{align}
    W_s = \frac{1}{2} (W + W^{'}), \\
    W_a = \frac{1}{2} (W - W^{'}),
\end{align}
where the signs $+$ and $-$ are the usual matrix addition and subtraction, and $W^{'}_s = \frac{1}{2} (W^{'} + W) = W_s$ is a symmetric weight matrix corresponding to the symmetric component $G_s$ and $W^{'}_a = \frac{1}{2}(W^{'} - W) = - W_a$ the anti-symmetric weight matrix that corresponds to the anti-symmetric component. Note that $W = W_s \bigoplus W_a$. This decomposition is unique, since it is the result of the projection of the weight matrix $W$ on the space of symmetric and anti-symmetric matrices, since $W_s$ and $W_a$ are respectively the closest symmetric and anti-symmetric matrices to $W$ in terms of the Frobenius norm; furthermore, the spaces of symmetric and anti-symmetric matrices are orthogonal with respect to standard matrix inner product.

In order to analyze the alterations in the topology of the network following the onset of seizure and during the seizure episode, we can use either $W_s$ or $W_a$. Since $W_s$ is a symmetric matrix, it can be used through a monotonically decreasing function, as it is the case with correlation and coherence matrices. However, $W_a$ being anti-symmetric, it needs to be transformed since a distance function cannot take negative values. Therefore, we propose to build the Rips-Vietoris filtration based on $|W_a|$, where the entries $|W_a|_{p, q} = \frac{1}{2} |W_{p, q} - W_{q, p}|$. Indeed, $|W_a|_{p, q}$ can be thought of as a distance, since it measures departure from symmetry. If $W$ is perfectly symmetrical, then $\forall p, q, (W_a)_{p, q} = 0$, and as $W$ departs from symmetry, i.e., $W_{p, q}$ changes and differs from $W_{q, p}$, $|W_a|_{p, q}$ becomes larger than zero. In the following section, we provide a brief review of persistent homology, before choosing to focus on $|W_a|$ to analyze the temporal evolution of the changes in directional asymmetry in brain connectivity during the seizure process.

\section{Overview of persistent homology and Vietoris-Rips filtration}
\label{sec:PH_VR_Filtration_Overview}

The goal of persistent homology is to provide computational tools that can distinguish between topological objects. It analyzes their topological features, such as connected components, holes, cavities, etc. In this paper we consider weighted networks, with weights corresponding to some loose notion of distance as defined by $|W_a|$. 

\begin{figure}
    \centering
    \includegraphics[width=.5\linewidth]{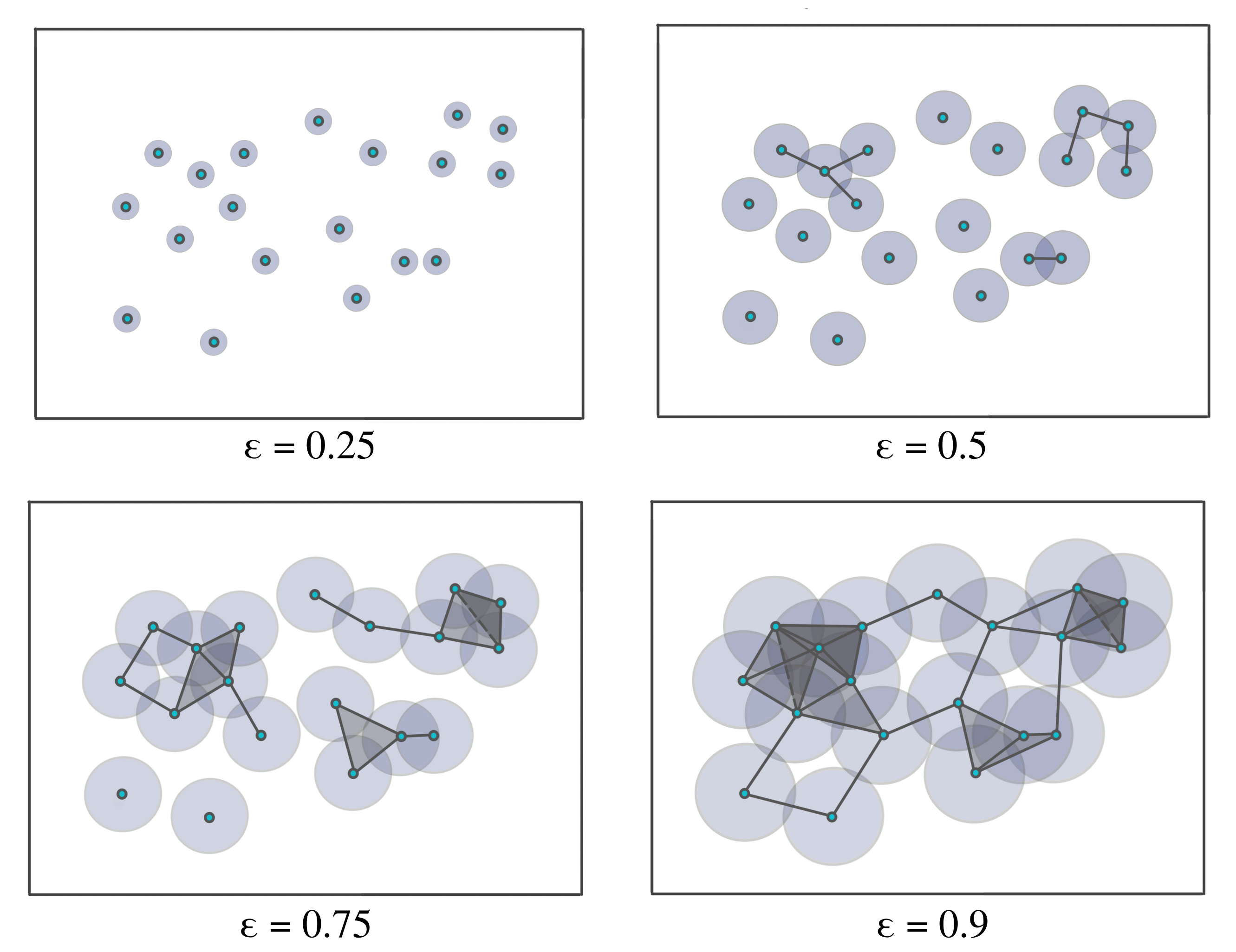}
    \caption{Example of a Vietoris-Rips filtration on a cloud of points. As the radius of the balls grows, there's birth and death of various topological features.}
    \label{fig:PH_example}
\end{figure}

\begin{figure}
    \centering
    \includegraphics[width=.5\linewidth]{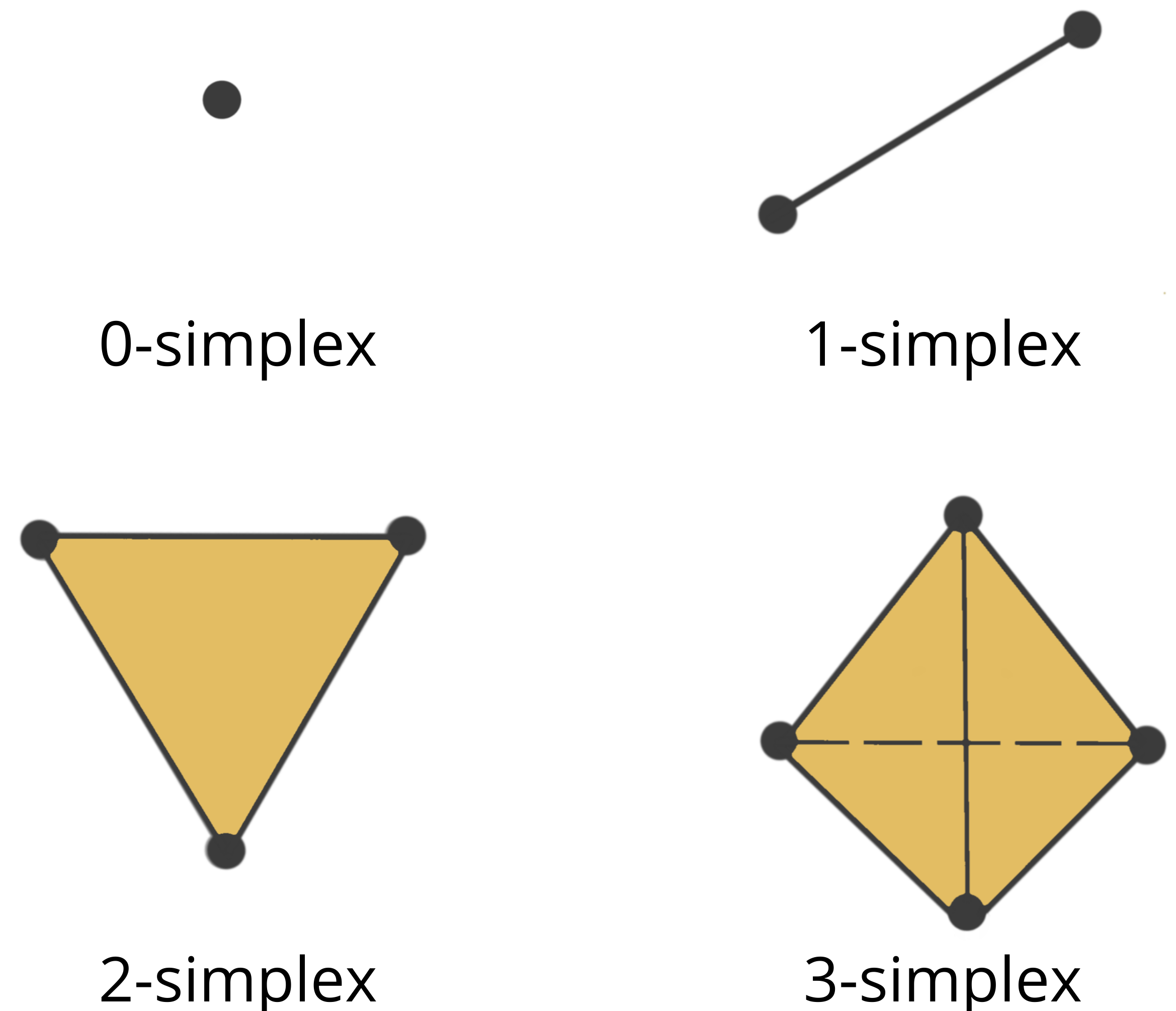}
    \caption{Examples of the four lowest dimensional simplices.}
    \label{fig:simplicial_complexes_example}
\end{figure}

\begin{figure}
    \centering
    \includegraphics[width=.6\linewidth]{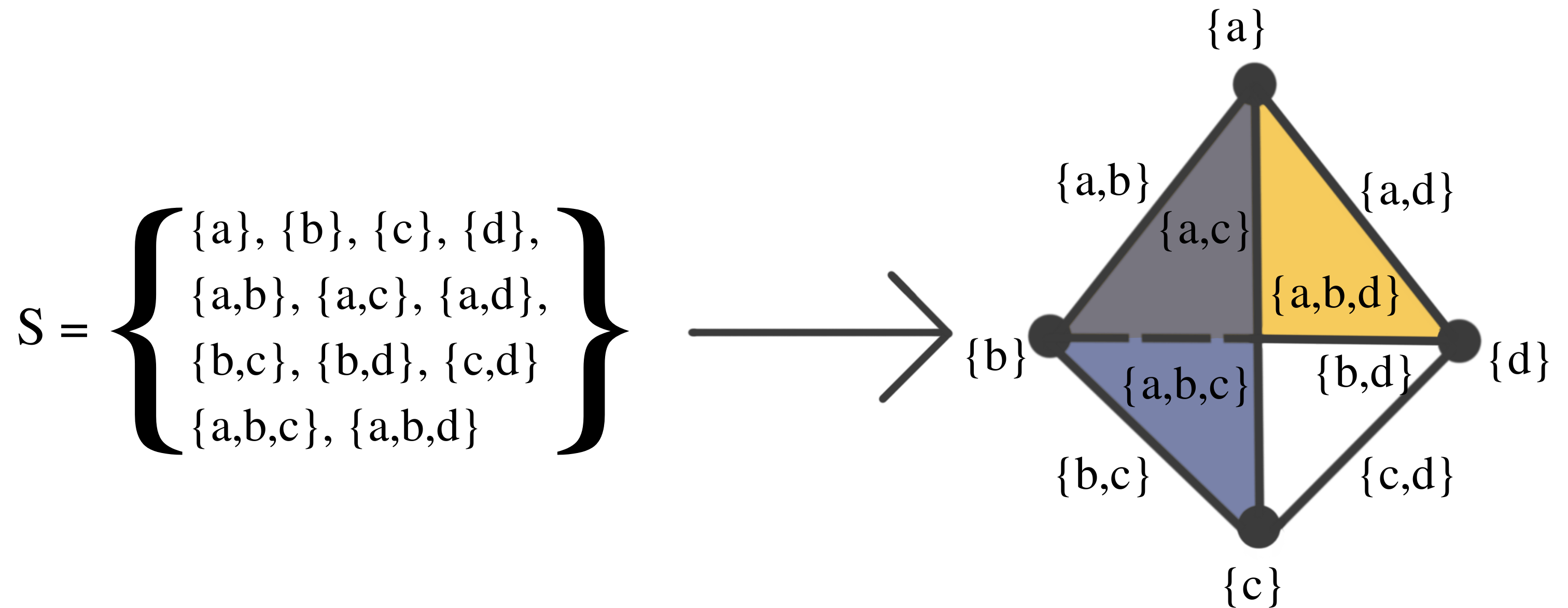}
    \caption{Example of a simplicial complex with four nodes, six edges and two faces. If $S$ is a simplicial complex, then every face of a simplex in $S$ must also be in $S$.}
    \label{fig:simplicial_complex_example}
\end{figure}

In order to analyze the shape of these networks, we need to build the homology of the data by looking at an increasing sequence of networks of neighboring data points at varying scales/distances, as seen in Figure \ref{fig:PH_example}. 
The goal of this approach is to analyze the characteristics of the geometrical patterns when they appear (birth) then disappear (death) for a wide range of scales, see \cite{BARCODES_FIRST}, \cite{EDELSBRUNNER_HARER}, \cite{TDA_EDELSBRUNNER} and \cite{BARCODES}. The Vietoris-Rips (VR) filtration, as presented above is constructed based on the notions of a simplex and simplicial complex, which can be thought of as a finite collection of sets that is closed under the subset relation, see Figures \ref{fig:simplicial_complexes_example} and \ref{fig:simplicial_complex_example}. Simplicial complexes are generally understood as higher dimensional generalizations of graphs. Simplicial complexes are meant to represent in an abstract form the shape of the data 
at various scales, in order to simplify and allow abstract manipulations.
Simplicial complexes as shown in Figure \ref{fig:simplicial_complex_example} can be very simple like a group of disconnected nodes, or more complex, e.g., a combination of pairs of connected nodes, triplets of triangles or any higher dimensional simplex. This notion of a simplicial complex can be understood as a generalization of the notion of networks, to include surfaces, volumes and higher dimensional objects.
Given this definition of simplicial complexes, we can formalize the notion of the VR filtration to be an increasing sequence of simplicial complexes. Let $X_p(t)$ be an observed time series of brain activity at location $p \in N$ and time $t \in \{1, ...T \}$. Therefore, we can think of a distance between brain channels at locations $p$ and $q$ to be their distance from symmetry, i.e., imbalance in information flow $D(p, q) = |W_{a, (pq)}|$. Using this measure, we construct the Vietoris-Rips filtration by connecting nodes that have a distance less or equal to some given threshold $\epsilon$, which results in the following filtration:
\begin{align}
    \mathcal{X}_{\epsilon_1} \subset \mathcal{X}_{\epsilon_2} \subset \cdots \subset \mathcal{X}_{\epsilon_n}, \label{eq:filtration}
\end{align}
where $0 < \epsilon_1 < \epsilon_2 < \cdots < \epsilon_{n-1} < \epsilon_n$ are the distance thresholds. Nodes within some given distance $\epsilon_i$ are connected to form different simplicial complexes, $\mathcal{X}_{\epsilon_1}$ is the first simplicial complex (single nodes) and $\mathcal{X}_{\epsilon_n}$ is the last simplicial complex (all nodes connected, i.e., a clique of size $d$, where $d$ is the number of nodes or dimension of the multivariate time series). Refer to \cite{HAUSMANN_RIPS_FILTRATION} for a review of how to build the VR filtrations.

\begin{figure}
    \centering
    \includegraphics[width=.4\linewidth]{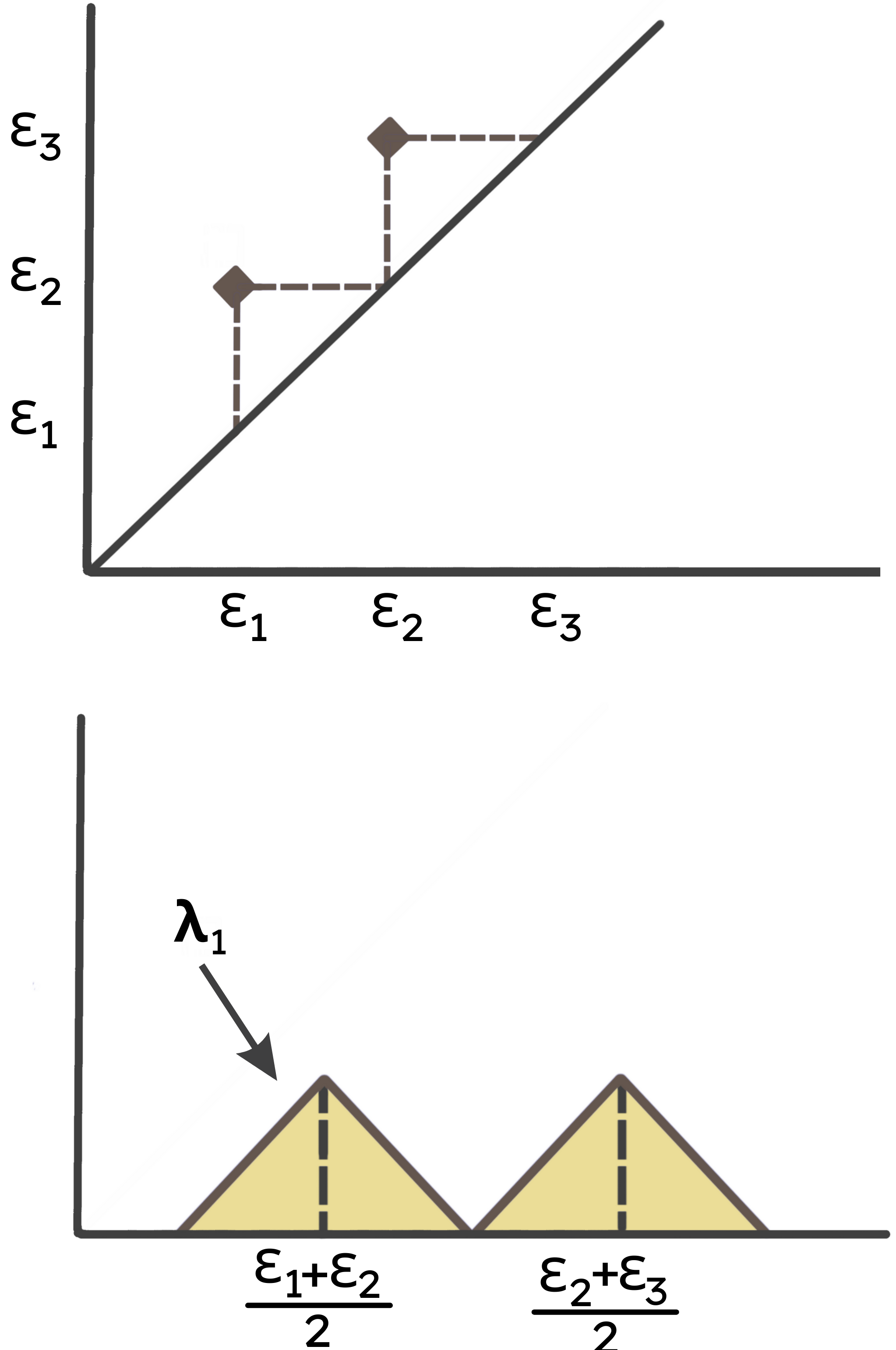}
    \caption{Example of a PD (Top) and a PL (Bottom). This Figure is inspired form the paper by \cite{TDA_GUIDEA}.}
    \label{fig:PD_PL_example_1}
\end{figure}

The VR filtration is a complex object. Therefore, in practice practitioners consider a topological summary that is known as the persistence diagram (PD), which is a diagram that represents the times of birth and death of the topological features in the VR filtration as seen in Figure \ref{fig:PD_PL_example_1}. Every birth-death pair is represented by a point in the diagram, e.g., ($\epsilon_1$, $\epsilon_2$) and ($\epsilon_2$, $\epsilon_3$). The points in the PD are colored based on the dimension of the feature they correspond to (e.g., one color for the connected components, another color for the cycles etc.).

Comparing PDs can be quite challenging and time consuming. Indeed, taking averages or computing distances, like the Bottleneck or Wasserstein distances  between persistence diagrams is time consuming as it is necessary to find point correspondence, see \cite{WASSERSTEIN_BOTELLNECK}. For these reasons, we prefer to analyze a simpler representation of the PD called the persistence landscape (PL), which is a simpler object (one dimensional function), as defined in \cite{PL_FIRST}. The use of PLs lends to a more rigorous statistical analysis that produce more easily interpretable results. As the PLs are functions of a real variable, it is easy to compute group averages and to derive confidence regions.

\section{Oriented TDA of seizure EEG}
\label{sec:data_analysis}

\begin{figure}
    \centering
    \includegraphics[width=.3\linewidth]{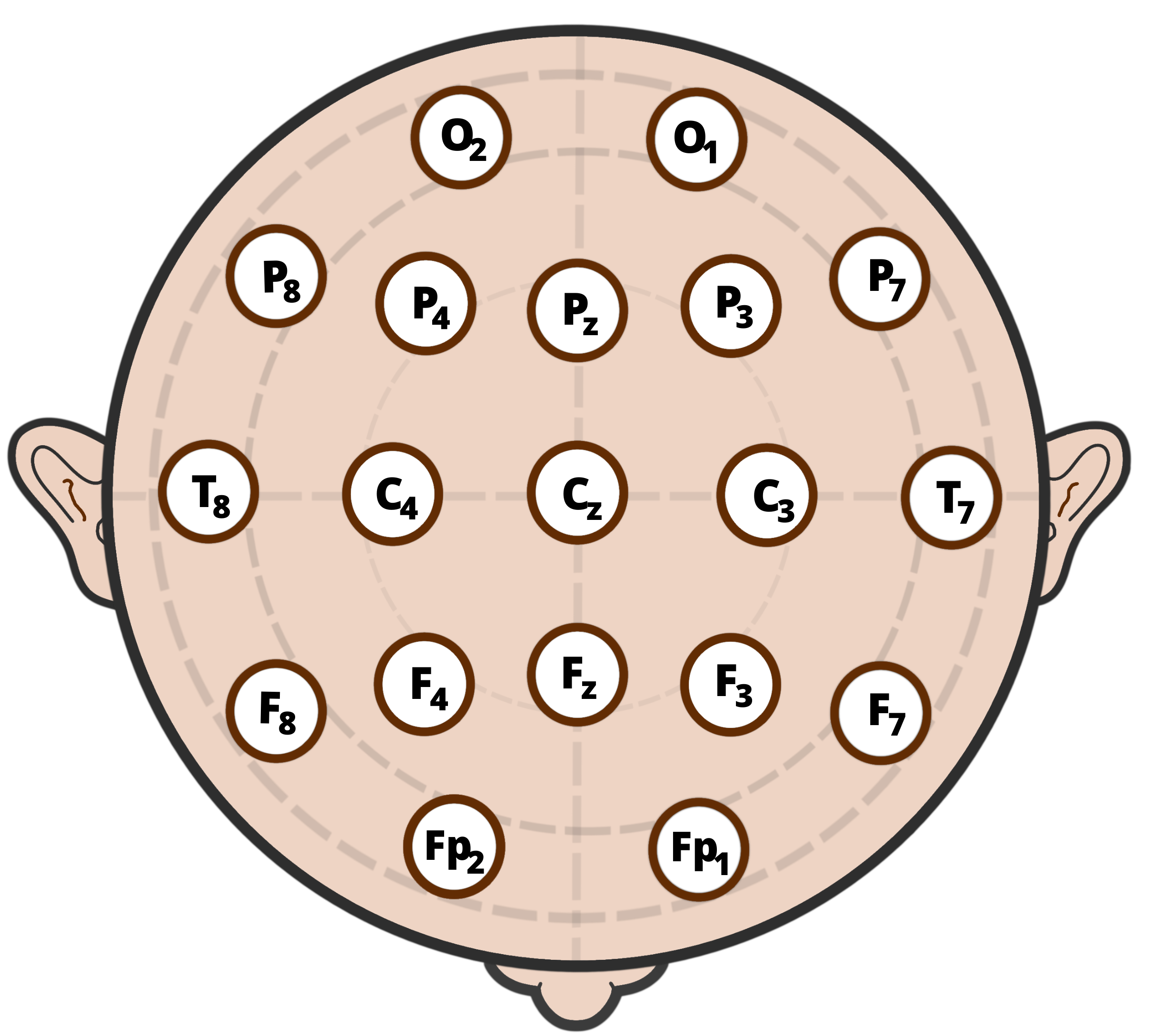}
    \caption{Differential electrodes positioning map on the scalp.}
    \label{fig:scalp_eeg_map}
\end{figure}

We investigate the topological features of the asymmetric brain network component based on the decomposition presented previously by contrasting the network before and after the epileptic seizure. The data was collected from an epileptic patient at the Epilepsy Disorder Laboratory at the University of Michigan (PI: Beth Malow, M.D.). With a sampling rate of 100Hz and observation period of 8 minutes, the following results are based on 19 scalp differential electrodes (no reference), see Figure \ref{fig:scalp_eeg_map}.

\begin{figure}
    \centering
    \includegraphics[width=\linewidth]{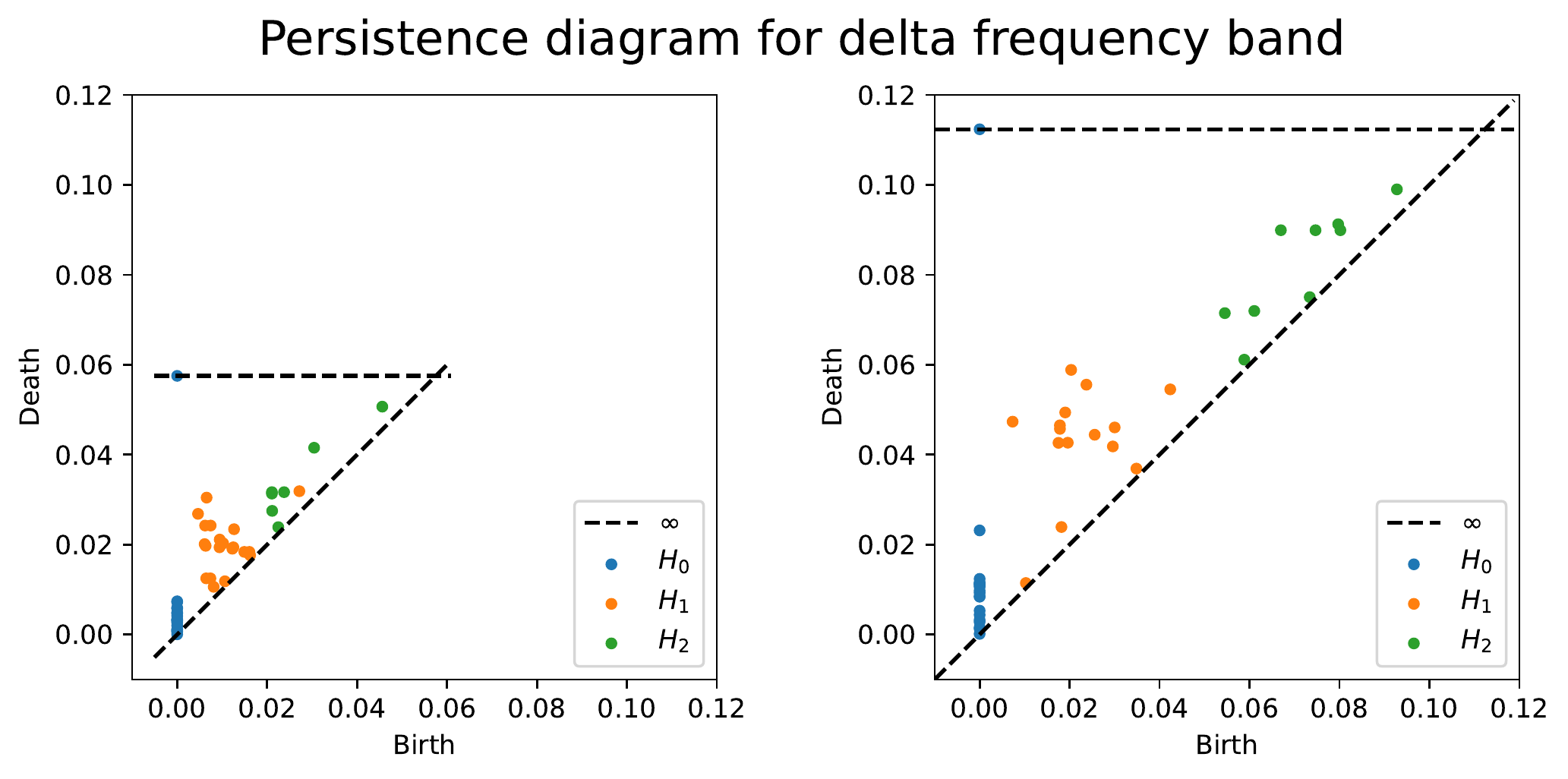}
    \caption{PD based on the anti-symmetric brain component for the delta (0 - 4Hz) frequency band. Pre-seizure (Left) and During-seizure (Right). It can be seen that zero- and one-dimensional features (blue and orange dots) are born around the same scale ($0.0$ and $0.01-0.02$) prior to and during the seizure and die at a much later scale during the seizure ($0.11$ and $0.06$) than prior to the seizure ($0.06$ and $0.03$). For the two-dimensional features (green dots), they seem to be born later and die later prior to and during the seizure.}
    \label{fig:PD_antisymmetric_delta}
\end{figure}

\begin{figure}
    \centering
    \includegraphics[width=\linewidth]{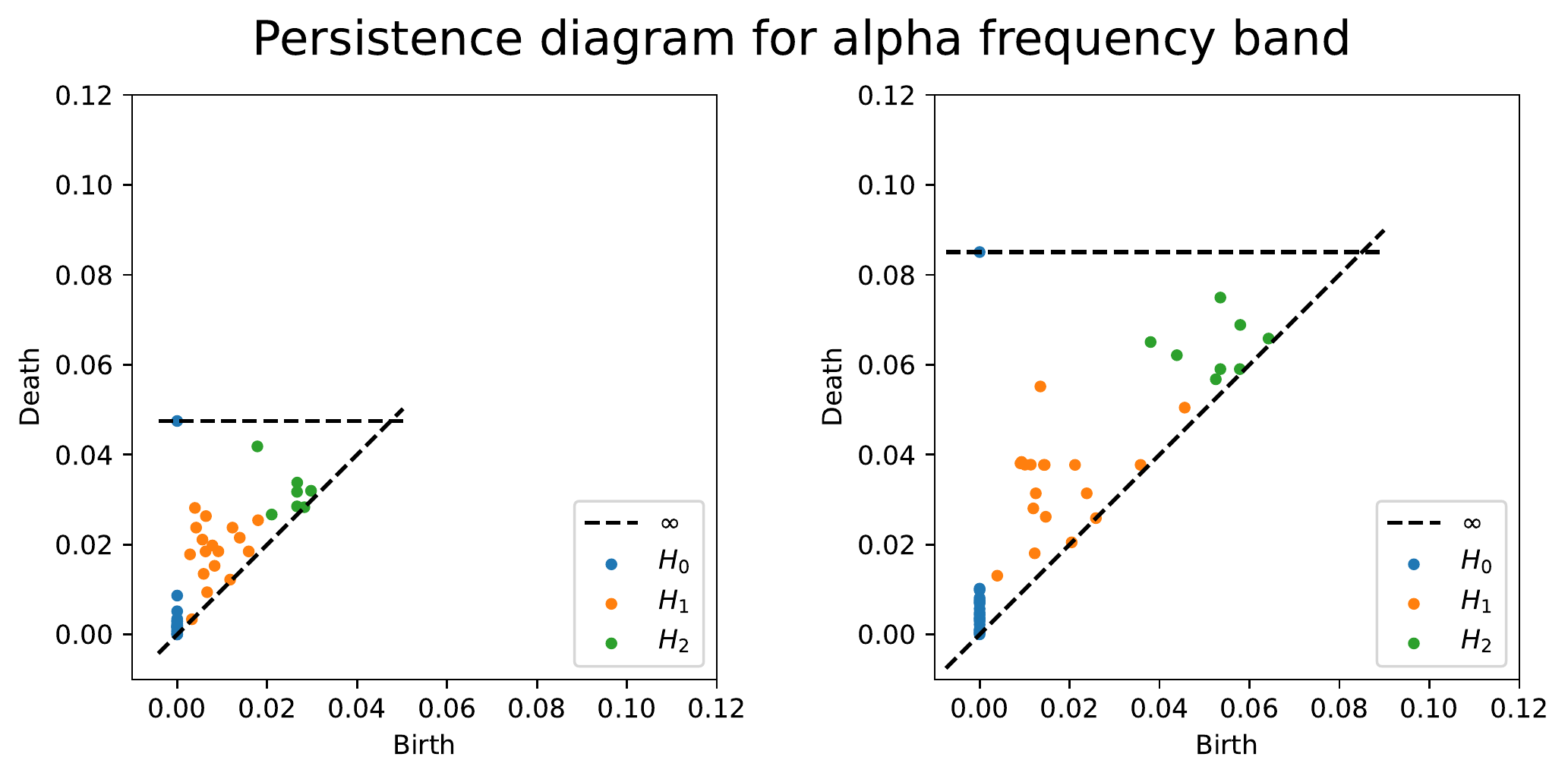}
    \caption{PD based on the anti-symmetric brain component for the alpha (8 - 12Hz) frequency band. Pre-seizure (Left) and During-seizure (Right). Similarly, it can be seen that zero- and one-dimensional features (blue and orange dots) are born around the same scale ($0.0$ and $0.01-0.02$) prior to and during the seizure and die at a much later scale during the seizure ($0.9$ and $0.06$) than prior to the seizure ($0.05$ and $0.03$). For the two-dimensional features (green dots), they seem to be born later and die later as well prior to and during the seizure.}
    \label{fig:PD_antisymmetric_alpha}
\end{figure}

\begin{figure}
    \centering
    \includegraphics[width=\linewidth]{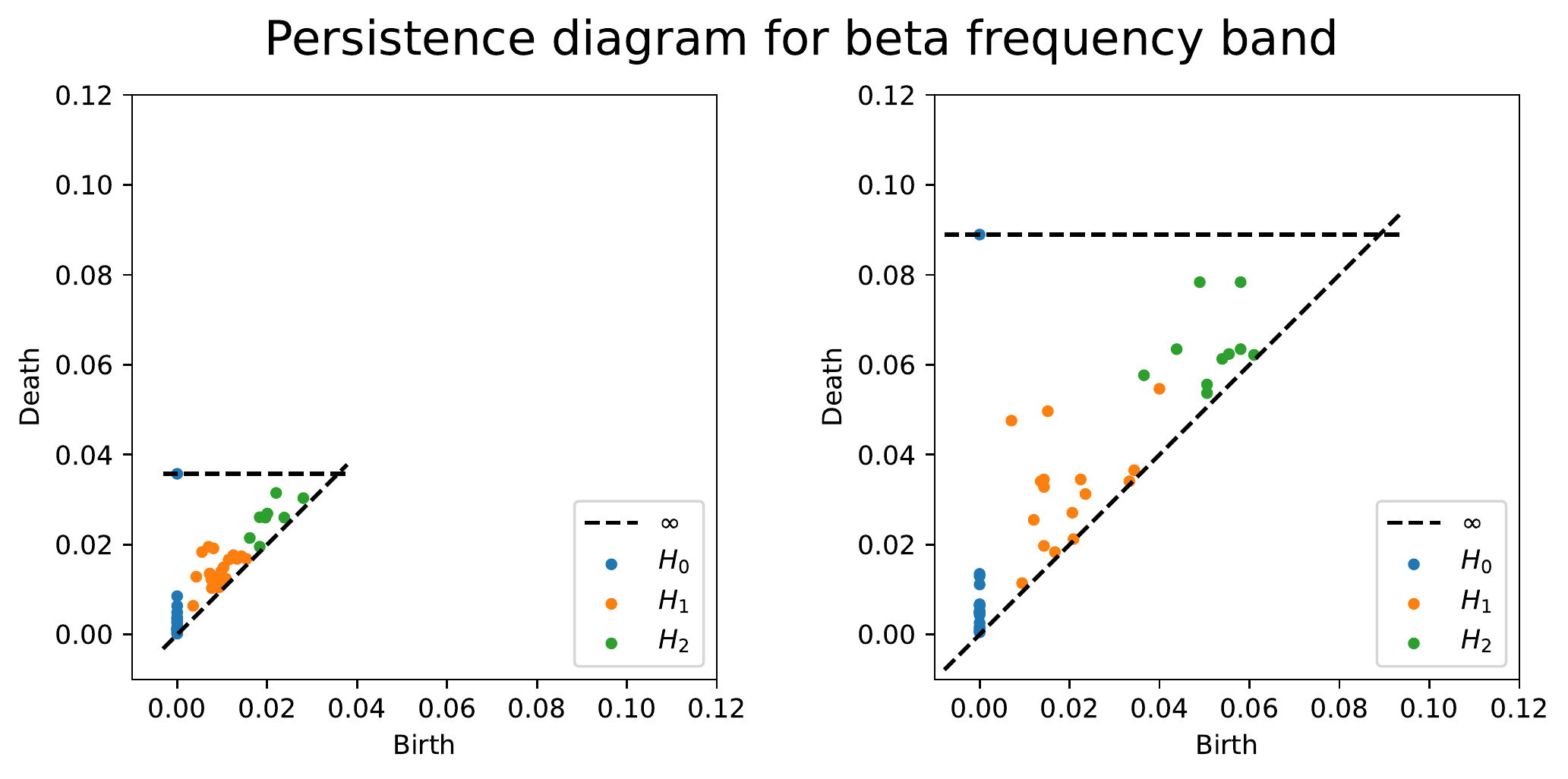}
    \caption{PD based on the anti-symmetric brain component for the beta (12 - 30Hz) frequency band. Pre-seizure (Left) and During-seizure (Right). The one-dimensional features (orange dots) seem to be born and die at a later scale respectively at ($0.01-0.04$)  and ($0.05$) during the seizure than prior to the seizure ($0.01-0.02$) and ($0.01-0.02$). Similarly, the two-dimensional features (green dots), seem to be born later and die later during the seizure than prior to the seizure.}
    \label{fig:PD_antisymmetric_beta}
\end{figure}

\begin{figure}
    \centering
    \includegraphics[width=\linewidth]{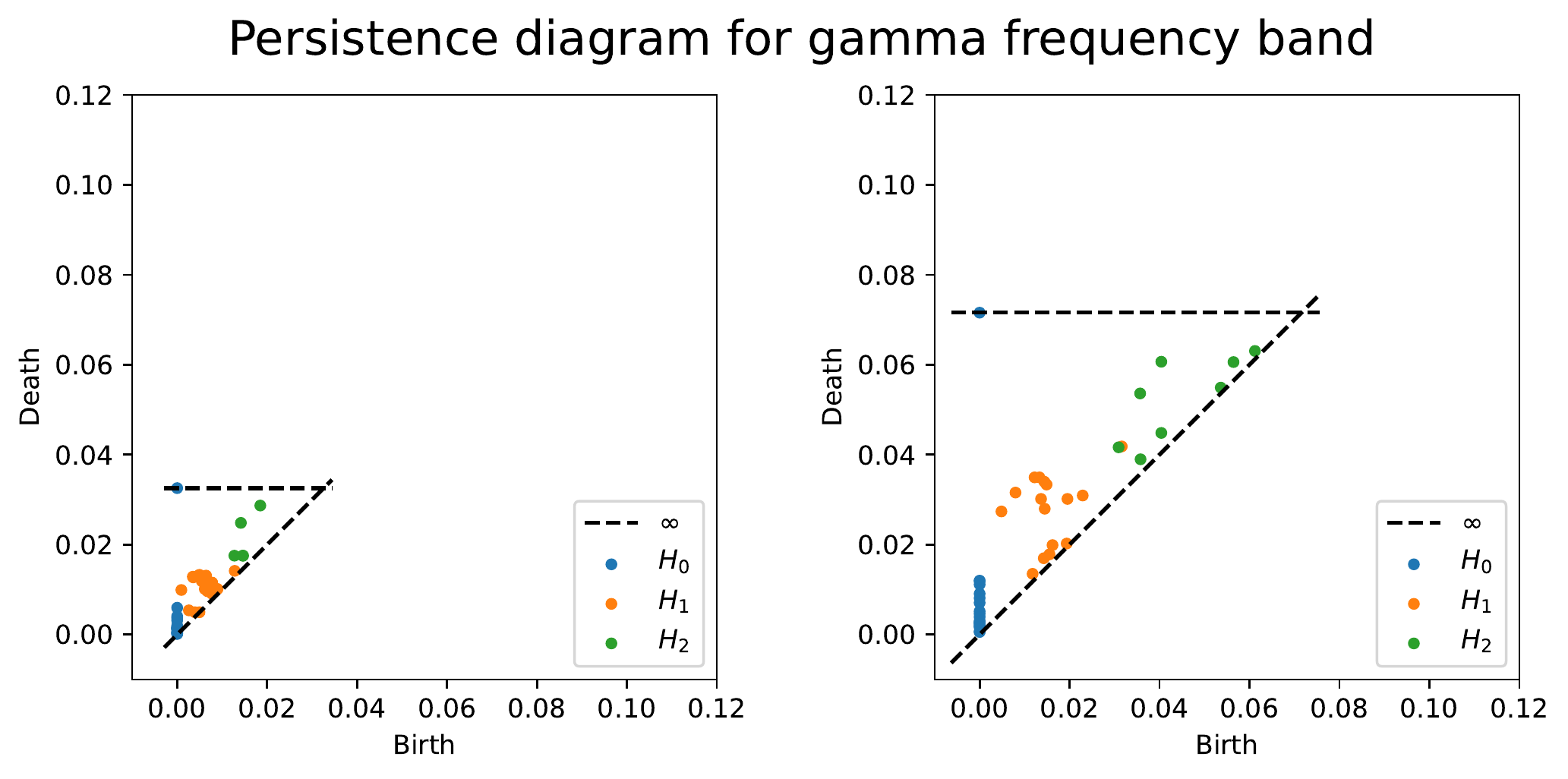}
    \caption{PD based on the anti-symmetric brain component for the gamma (30 - 50Hz) frequency band. Pre-seizure (Left) and During-seizure (Right). The zero- and one-dimensional features (blue and orange dots) seem to be born at the same time ($0.0$) and ($0.01-0.02$) but die at a later scale ($0.04$) during the seizure than prior to the seizure ($0.03$). Similarly, the two-dimensional features (green dots), seem to be born later and die later during the seizure than prior to the seizure.}
    \label{fig:PD_antisymmetric_gamma}
\end{figure}

To analyze this data set on the pre-seizure an seizure onset, we start by fitting a VAR model with dimension $d=19$ (number of observed time series components) and order $K=5$ to the pre-seizure part and on the seizure onset. Connectivity between all pairs of channels was assessed using PDC, for the delta (0 - 4Hz), alpha (8 - 12Hz), beta (12 - 30Hz) and gamma (30 - 50Hz) frequency bands. After we decompose each network into its symmetric and anti-symmetric components, then we build the VR filtration based on $|W_a^{\Omega}|$, where $\Omega$ stands for the frequency band of interest, delta, alpha, beta, gamma etc. We report the PDs in Figures \ref{fig:PD_antisymmetric_delta}-\ref{fig:PD_antisymmetric_gamma}.

From the above PDs, we notice that the 1-dimensional features persist more during the seizure than prior to the seizure mainly for delta, alpha and beta frequency bands. However, even if the 2-dimensional features seem to persist more during the seizure than prior to the seizure, the variability seems larger which makes conclusions uncertain.

In this context of epileptic seizure, the appearance of 1-dimensional features means that the epileptic seizure tends to induce an asymmetric circular flow of information within the brain. The appearance of 0-dimensional features indicates the presence of segregated regions where the flow of information is asymmetric. The scale at which these features appear indicates the importance or magnitude of the asymmetry, similarly the scale at which these features die indicates the maximum magnitude that the asymmetry within such features might reach.

The fact that these features appear at higher scales indicates that the epileptic seizure induces a large asymmetry in the flow of information within brain regions. Furthermore, since these features persist for larger scales,  it means that the epileptic seizure induces a non-homogeneous asymmetry throughout the brain regions.

\section{Conclusion}
\label{sec:conclusion}

Topological data analysis has been built upon the notion of distance, which is symmetric by definition. When applied to brain connectivity networks, current TDA tools encounter serious limitations as they can only analyze brain networks based on symmetric dependence measures such as correlations or coherence. We present a new approach to analyze asymmetric brain networks using more general measures of dependence. Our approach relies on network decomposition, using weight matrix projections onto symmetric and anti-symmetric matrix spaces. Using PDC, our approach is able to provide new insight into the topological alterations induced by an epileptic seizure. Our preliminary analysis provides evidence that epileptic seizure induces asymmetric flow of information across the brain, and that the induced asymmetry is non-homogeneous throughout brain regions. Even if we can detect seizure induced asymmetry in the flow of information within the brain, our approach cannot detect the origin of this asymmetry. In future work we plan to combine our oriented TDA approach with graph theoretic measures to detect the nature of the seizure as well as the regions at the origin of the seizure.

\nocite{*}
\bibliographystyle{spstat}
\bibliography{sample}

\begin{thebibliography}{44}
\providecommand{\natexlab}[1]{#1}
\providecommand{\url}[1]{{#1}}
\providecommand{\urlprefix}{URL }
\expandafter\ifx\csname urlstyle\endcsname\relax
  \providecommand{\doi}[1]{DOI~\discretionary{}{}{}#1}\else
  \providecommand{\doi}{DOI~\discretionary{}{}{}\begingroup
  \urlstyle{rm}\Url}\fi
\providecommand{\eprint}[2][]{\url{#2}}

\bibitem{WASSERSTEIN_BOTELLNECK}
\textsc{Agami, S.} (2021). Comparison of persistence diagrams. {\it
  Communications in Statistics - Simulation and Computation} \textbf{0} 1--14.

\bibitem{AIC}
\textsc{Akaike, H.} (1974). A new look at the statistical model identification.
  {\it IEEE Transactions on Automatic Control} \textbf{19} 716--723.

\bibitem{Bacala_Connectivity}
\textsc{Baccala, L. and Sameshima, K.} (2001). Partial directed coherence: A
  new concept in neural structure determination. {\it Biological Cybernetics}
  \textbf{84} 463--474.

\bibitem{BRAIN_NETWORKS_HEALTHY_DISEASE}
\textsc{Bassett, D. and Bullmore, E.} (2009). Human brain networks in health
  and disease. {\it Current opinion in neurology} \textbf{22} 340--347.

\bibitem{EPILEPSY_INTRO}
\textsc{Bromfield, E.~B., Cavazos, J.~E. and Sirven, J.~I.} (2006). {\em An
  Introduction to Epilepsy}. American Epilepsy Society.

\bibitem{PL_FIRST}
\textsc{Bubenik, P.} (2015). Statistical topological data analysis using
  persistence landscapes. {\it Journal of Machine Learning Research}
  \textbf{16} 77–102.

\bibitem{PL_SECOND}
\textsc{Bubenik, P.} (2020). The persistence landscape and some of its
  properties. {\it Topological Data Analysis: The Abel Symposium} \textbf{15}
  97–117.

\bibitem{GRAPHS_BRAIN_DATA}
\textsc{Bullmore, E. and Sporns, O.} (2009). Complex brain networks: Graph
  theoretical analysis of structural and functional systems. {\it Nature
  reviews Neuroscience} \textbf{10} 186--98.

\bibitem{BATESIAN_MODELING_MULTIVARIATE_TS}
\textsc{Cadonna, A., Kottas, A. and Prado, R.} (2019). Bayesian spectral
  modeling for multiple time series. {\it Journal of the American Statistical
  Association} \textbf{114} 1838--1853.

\bibitem{TDA_GUNNAR}
\textsc{Carlsson, G.} (2009). Topology and data. {\it Bulletin of the American
  Mathematical Society} \textbf{46} 255--308.

\bibitem{BARCODES_FIRST}
\textsc{Carlsson, G., Zomorodian, A., Collins, A. and Guibas, L.} (2004).
  Persistence barcodes for shapes.Association for Computing Machinery. p
  124–135.

\bibitem{PD_STABILITY}
\textsc{Cohen-Steiner, D., Edelsbrunner, H. and Harer, J.} (2007). Stability of
  persistence diagrams. {\it Discrete and computational geometry} \textbf{37}
  103--120.

\bibitem{EDELSBRUNNER_HARER}
\textsc{Edelsbrunner, H. and Harer, J.} (2008). Persistent homology—a survey.
  {\it Discrete and Computational Geometry} \textbf{453} 257--282.

\bibitem{TDA_EDELSBRUNNER}
\textsc{Edelsbrunner, H., Letscher, D. and Zomorodian, A.} (2002). Topological
  persistence and simplification. {\it Discrete \& Computational Geometry}
  \textbf{28} 511–533.

\bibitem{TDA_MULTIVARIATE_TS_ANASS}
\textsc{El-Yaagoubi, A., Chung, M.~K. and Ombao, H.} (2022). Topological data
  analysis for multivariate time series data. {\it arXiv} \textbf{0} 0.

\bibitem{FUNCTIONAL_CON_1}
\textsc{Fiecas, M., Ombao, H., Linkletter, C., Thompson, W. and Sanes, J.}
  (2010). Functional connectivity: shrinkage estimation and randomization test.
  {\it Neuroimage} \textbf{4} 15--49.

\bibitem{FUNCTIONAL_CON_2}
\textsc{Fiecas, M., Ombao, H., van Lunen, D., Baumgartner, R., Coimbra, A. and
  Feng, D.} (2013). Quantifying temporal correlations: a test-retest evaluation
  of functional connectivity in resting-state fmri. {\it Neuroimage}
  \textbf{65} 231--41.

\bibitem{TDA_TSA}
\textsc{Gholizadeh, S. and Zadrozny, W.} (2018). A short survey of topological
  data analysis in time series and systems analysis. {\it ArXiv}
  \textbf{abs/1809.10745}.

\bibitem{BARCODES}
\textsc{Ghrist, R.} (2008). Barcodes: The persistent topology of data. {\it
  Bulletin of the American Mathematical Society} \textbf{45} 61--75.

\bibitem{TDA_GUIDEA}
\textsc{Gidea, M. and Katz, Y.} (2018). Topological data analysis of financial
  time series: Landscapes of crashes. {\it Physica A: Statistical Mechanics and
  its Applications} \textbf{491} 820--834.

\bibitem{GIUTI_CLIQUE_TOPOLOGY}
\textsc{Giusti, C., Pastalkova, E., Curto, C. and Itskov, V.} (2015). Clique
  topology reveals intrinsic geometric structure in neural correlations. {\it
  Proceedings of the National Academy of Sciences} \textbf{112} 13455--13460.

\bibitem{VAR_1}
\textsc{Gorrostieta, C., Ombao, H., Bedard, P. and Sanes, J.} (2012).
  Investigating stimulus-induced changes in connectivity using mixed effects
  vector autoregressive models. {\it NeuroImage} \textbf{59} 3347--3355.

\bibitem{VAR_2}
\textsc{Gorrostieta, C., Fiecas, M., H~Ombao, E.~B. and Cramer, S.} (2013).
  Hierarchical vector auto-regressive models and their applications to
  multi-subject effective connectivity. {\it Frontiers in Computational
  Neuroscience} \textbf{159} 1--11.

\bibitem{AR_MIXTURES_BRAIN_SIGNALS}
\textsc{Granados-Garcia, G., Fiecas, M., Babak, S., Fortin, N.~J. and Ombao,
  H.} (2021). Brain waves analysis via a non-parametric bayesian mixture of
  autoregressive kernels. {\it Computational Statistics \& Data Analysis}
  \textbf{174} 107409.

\bibitem{GRANGER}
\textsc{Granger, C. W.~J.} (1969). Investigating causal relations by
  econometric models and cross-spectral methods. {\it Econometrica} \textbf{37}
  424--438.

\bibitem{HAUSMANN_RIPS_FILTRATION}
\textsc{Hausmann, J.-C.} (2016). {\em On the Vietoris-Rips complexes and a
  Cohomology Theory for metric spaces}. , vol 138Princeton University Press.

\bibitem{SMALL_WORLD_BRAIN}
\textsc{Hilgetag, C. and Goulas, A.} (2015). Is the brain really a small-world
  network? {\it Brain structure \& function} \textbf{221} 2361–2366.

\bibitem{LLN_CLT}
\textsc{Hoffmann-Jorgensen, J. and Pisier, G.} (1976). The law of large numbers
  and the central limit theorem in banach spaces. {\it The Annals of
  Probability} \textbf{4} 587--599.

\bibitem{VAR_4}
\textsc{Hu, L., Fortin, N. and Ombao, H.} (2019). Vector autoregressive models
  for multivariate brain signals. {\it Statistics in the Biosciences}
  \textbf{11} 91--126.

\bibitem{VAR_3}
\textsc{Kirch, C., Muhsal, B. and Ombao, H.} (2015). Detection of changes in
  multivariate time series with application to eeg data. {\it Journal of the
  American Statistical Association} \textbf{110} 1197--1216.

\bibitem{PROBABILITY_BANACH_SPACES}
\textsc{Ledoux, M. and Talagrand, M.} (2011). {\em Probability in Banach
  Spaces. Classics in Mathematics}. Springer-Verlag, Berlin.

\bibitem{FUNCTIONAL_CONNECTIVITY_ALZHEIMER}
\textsc{Leuchter, A.~F., Newton, T.~F., Cook, I.~A., Walter, D.~O.,
  Rosenberg-Thompson, S. and Lachenbruch, P.~A.} (1992). Changes in brain
  functional connectivity in alzheimer-type and multi-infarct dementia. {\it
  Brain} \textbf{115} 1543--1561.

\bibitem{LUTKEPOHL}
\textsc{Lütkepohl, H.} (1991). {\em New Introduction to Multiple Time Series
  Analysis}. Springer.

\bibitem{MERKULOV_ALG_TOP}
\textsc{Merkulov, S.} (2003). Algebraic topology. {\it Proceedings of the
  Edinburgh Mathematical Society} \textbf{46}.

\bibitem{SMALL_WORLD_NETWORKS}
\textsc{Muldoon, S., Bridgeford, E. and Bassett, D.} (2016). Small-world
  propensity and weighted brain networks. {\it Scientific Reports} \textbf{6}
  22057.

\bibitem{MUNKRES_ALG_TOP}
\textsc{Munkres, J.~R.} (1984). {\em Elements of Algebraic Topology}. Addison
  Wesley Publishing Company.

\bibitem{SPECTRAL_DEPENDENCE}
\textsc{Ombao, H. and Pinto, M.} (2021). Spectral dependence.
  \eprint{2103.17240}

\bibitem{ONBAO_BELLEGEM}
\textsc{Ombao, H. and Van~Bellegem, S.} (2008). Evolutionary coherence of
  nonstationary signals. {\it IEEE Transactions on Signal Processing}
  \textbf{56} 2259--2266.

\bibitem{BRAIN_NETWORKS_ORGANIZATION}
\textsc{Pessoa, L.} (2014). Understanding brain networks and brain
  organization. {\it Physics of Life Reviews} \textbf{11} 400--435.

\bibitem{AR_MIXTURE_MODELS}
\textsc{Prado, R.} (2013). Sequential estimation of mixtures of structured
  autoregressive models. {\it Computational Statistics \& Data Analysis}
  \textbf{58} 58--70.

\bibitem{BIC}
\textsc{Schwarz, G.} (1978). Estimating the dimension of a model. {\it The
  Annals of Statistics} \textbf{6} 461--464.

\bibitem{TSA_SHUMWAY_STOFFER}
\textsc{Shumway, R.~H. and Stoffer, D.~S.} (2005). {\em Time Series Analysis
  and Its Applications}. Springer-Verlag.

\bibitem{STRUCTURE_FUNCTION_BRAIN_NETWORKS}
\textsc{Sporns, O.} (2013). Structure and function of complex brain networks.
  {\it Dialogues in Clinical Neuroscience} \textbf{15} 247--262.

\bibitem{BRAIN_SIGNALS_LOW_EMBEDDING}
\textsc{Wang, Y., Ting, C.-M., Gao, X. and Ombao, H.} (2019). Exploratory
  analysis of brain signals through low dimensional embedding 997--1002.

\end{thebibliography}
\end{document}